\documentclass[11pt]{article}
\makeatletter
\def\ps@pprintTitle{%
 \let\@oddhead\@empty
 \let\@evenhead\@empty
 \def\@oddfoot{}%
 \let\@evenfoot\@oddfoot}
\makeatother
\usepackage{mathrsfs} 
\usepackage{hyperref,graphicx}
\usepackage[normalem]{ulem}
\usepackage[margin=1.0in]{geometry}
\usepackage{multirow}
\usepackage{booktabs, threeparttable}
\usepackage{subfigure}
\usepackage{physics}
\usepackage{braket}
\usepackage{dsfont}
\usepackage{cite}
\usepackage[cmintegrals,varg]{newtxmath} 

\allowdisplaybreaks[2]
\usepackage{amsfonts}
\usepackage{amsmath,bm}
\usepackage{booktabs, threeparttable}
\usepackage{array}
\newcolumntype{L}[1]{>{\raggedright\arraybackslash}p{#1}}
\usepackage{color}

\usepackage{authblk}
\title{Thermal response functions and second sound in graphene}
\author[1]{Antonio Martinez Margolles}
\author[1,2]{Patrick K. Schelling \thanks{Corresponding author, patrick.schelling@ucf.edu}}
\affil[1]{Department of Physics, University of Central Florida, Orlando, FL 32816-2385, USA}
\affil[2]{Advanced Materials Processing and Analysis Center, University of Central Florida, Orlando, FL 32816-2385, USA}
\begin{document}
\maketitle

\begin{abstract}
The propagation of second sound, and more broadly the ballistic transport of heat, is of central importance in heat dissipation from electronic devices at very short length and time scales. Recently, we have developed thermal-response functions appropriate for elucidating physics beyond the diffusive regime, including time-dependent sources and wave-like heat propagation.
The methods are applied to graphene simulated using molecular-dynamics (MD) with empirical potentials. The simulations predict a strong oscillatory transport at $T=300$K for length scales equal to $L=68.1$nm and below. It is shown that at these temperatures and scales, the lifetime of the oscillatory transport is determined largely by wave coherence connected to the phonon band structure. While most BTE theories for second sound neglect this effect, and may not be suitable at very short length scales, they nevertheless are accurate for describing perturbations at longer length scales. Calculations
using the linearized BTE (LBTE) are also presented, along with analysis of second sound. This
approach results in significantly longer lifetimes for second sound in comparison to our MD simulation results. Predictions for the response due to time-dependent sources are also presented, including insight into how time-dependent experiments might probe the spectra associated with second sound. Results are discussed in relation to recent experiments on graphite. 
 \end{abstract}


\maketitle

\section{Introduction}
Recent experiments using transient-thermal grating (TTG) techniques have demonstrated the presence of second-sound propagation in graphite \cite{huberman_2019,Ding_2022} at temperatures
over $200$K. The relevance of this transport mechanism to applications is established by the fact that it was observed for grating periods of about 2$\mu$m.  Moreover, theoretical calculations of graphene have established that boundary scattering effects are dominant at scales $\sim 1\mu$m \cite{Bonini_2012} which substantiates the relevance of ballistic transport to potential device applications. Despite these facts, the vast majority of theoretical calculations are focused on computing thermal
conductivity $\kappa$, which presume the relevance of the heat-diffusion equation, and hence are not relevant for transport via second sound. More broadly, most approaches are
insufficient for a description of time-dependent heat sources when Fourier's law does not apply.

The phenomenological description of second sound has been based on the hyperbolic heat equation for the evolution of the temperature field $T(\bm{r},t)$,
\begin{equation}
\label{hheq}
{\partial^{2} T(\bm{r},t)\over \partial t^{2}} 
+{1 \over \tau_{ss}} {\partial T(\bm{r},t) \over \partial t}
-v_{ss}^{2} \nabla^{2} T(\bm{r},t) =0   \text{.}
\end{equation}
This equation, written in Fourier space, with $T_{\bm{q}}(t)$ representing a deviation from equilibrium with wave vector $\bm{q}$, is described by the damped oscillator equation,
\begin{equation}
\label{hheq2}
{d^{2} T_{\bm{q}}\over d t^{2}} 
+{1 \over \tau_{ss}} {d T_{\bm{q}} \over d t}
+q^{2}v_{ss}^{2} T_{\bm{q}} =0   \text{,}
\end{equation}
in which $q^{2} = \bm{q}\cdot\bm{q}$.
Hence, perhaps more specifically, oscillations in the temperature field can be observed when  
$ \tau_{ss}> {1 \over 2q v_{ss}}$. Because $\tau_{ss}$ and $v_{ss}$ represent the lifetime and
propagation velocity of temperature waves, this is simply stating that $\tau_{ss}$ must be large enough
to encompass at least one period of oscillation for wave behavior to be observed. If $\tau_{ss}< {1 \over 2q v_{ss}}$, a temperature
deviation $T_{\bm{q}}(t)$ will simply exponentially decay with time. Finally, if $\tau_{ss} \ll {1 \over 2q v_{ss}}$, eventually Fourier's
law behavior is expected to apply. It is standard to use solutions to the BTE to predict values for  $v_{ss}$ and $\tau_{ss}$\cite{ENZ1968114,Hardy:1970aa}. More recently, theoretical predictions have been grounded in first-principles calculations typically using Density-Functional Theory (DFT) \cite{Lee_2015,Cepellotti_2015,Lee:2017aa,Luo:2019aa,Shang:2022aa}.

In this paper, we apply our recently-developed \cite{Fernando_2020,Bohm:2022aa,Schelling:2025aa} approach using classical molecular-dynamics (MD) simulation  to compute the thermal-response functions for graphene at $T=300$K. The advantage of this approach is that it makes predictions without any assumptions  about whether transport is diffusive or ballistic, and moreover is directly comparable to experiments using TTG techniques. In previous works, response functions were able to identify the presence of second sound in one-dimensional chains \cite{Bohm:2022aa} and hBN monolayers \cite{Schelling:2025aa}. Here we demonstrate
the presence of oscillatory heat transport in graphene at $T=300$K over length scales up to $L=68.1$nm. The phenomenological expression in Eq. \ref{hheq2} is compared to the MD simulations to obtain values for $\tau_{ss}$ and $v_{ss}$.  This is then compared to predictions based on density-functional theory (DFT) calculations and theoretical expressions for second-sound based on Ref. \cite{Cepellotti_2015}. It will be
shown, in contrast to the assumptions made by BTE-based theories of second sound, that $\tau_{ss}$ strongly depends on the wave vector $\bm{q}$ of the perturbation. Moreover, it will be demonstrated that the primary mechanism for the decay in the oscillatory signal at $T=300$K at these length scales is the loss of phase coherence due to normal-mode dispersion rather than 
anharmonic scattering.  By contrast, simulations at $T=10K$, while still demonstrating the importance of phonon dispersion, also shows sharp spectral features and more complex behavior. An external heat pulse like those generated in a TTG experiment generates phase coherence between normal modes which is lost as different modes in the  wave propagate at different velocities. Comparison to experimental TTG techniques applied to graphite appear to follow this qualitative picture \cite{Ding_2022}. At short length scales (i.e. the grating period), this mechanism leads to values for $\tau_{ss}$ which are much smaller than those obtained from BTE theories. However, at longer length scales, we surmise that BTE theories provide an accurate description. Finally, we examine the physics of time-dependent perturbations, which is used to develop insight into how time-dependent sources could be used to probe the spectra of second sound.

We note here that the term ``second sound'' is often used broadly to indicate the presence of oscillatory heat transport. However, the general paradigm for ``second sound'' is that it corresponds to oscillatory heat transport with corresponding collective, hydrodynamic phonon transport which occurs due to the predominance of normal phonon scattering.  Including this requirement generates a more restrictive definition of ``second sound''. In the following, we use this more restrictive definition, and hence we can only hypothesize that the oscillatory transport observed in our simulations might be consistent with second sound. However, at present we are unable to establish displaced phonon distributions and therefore it is possible that we are observing oscillatory transport without underlying hydrodynamic behavior. Yet it is important to note that it is generally understood that the conditions for predominance of normal scattering exists in graphene at low temperatures.

\section{Approach: Classical MD}

The interactions to describe graphene are taken from the optimized Tersoff potential reported in Ref. \cite{Lindsay:2010vb}. 
As with our previous study \cite{Schelling:2025aa}, the expressions for heat current due to Fan and coworkers \cite{Fan:2015tj} were used. The thermal conductivity $\kappa$ was
computed according to the Green-Kubo expression,
\begin{equation} \label{GK}
\kappa(\tau ) = {\Omega  \over 3 k_{B}T^{2}} \int_{0}^{\tau} \langle \bm{J}(t)\cdot \bm{J}(0) \rangle dt
\end{equation}
with the thermal conductivity formally given by taking the limit $\kappa =  \lim_{\tau \rightarrow \infty} \kappa(\tau)$. To define the system volume $\Omega$ for the 
two-dimensional system, the layer spacing in graphite $c=3.35 \AA$ was used. For the in-plane directions, the lattice parameters were chosen to result in $T=0$K C-C bond length of $1.44$\AA, which is fairly close to the zero-stress result in Ref. \cite{Lindsay:2010vb}, but slightly expanded from the experimental value $\sim 1.42\AA$. All simulations were conducted at constant volume, and hence at $T=300$K, the system is under tensile stress since thermal contraction cannot occur. Accurate calculation of $\kappa$ requires 
thermal equilibration of the simulated system and large amounts of simulation data for statistical averaging. We used an ensemble of 640 
independent calculations each with 0.262 ns of simulation data within the constant energy and volume ensemble to compute the time-averaged current-current correlation function. Hence, the total amount of simulation time used for averaging 
was 167.7 ns. The time step for integration of the equations of motion was $0.146$ fs which was small enough to demonstrate excellent energy conservation. Finally, the system 
was chosen to have dimensions $320 \times 32$ cells along the $\bm{a}_{1}$ and $\bm{a}_{2}$ lattice vectors. Hence the system size comprised $N=20,480$ atoms. The supercell
was chosen in this way such that coherent wave-like transport could be examined over long length scales without excessive computational cost.  The supercell used
was $L=68.1$ nm along the long direction. While we have not explored dependence of the results on transverse dimensions, we note here that the computed results at $T=300$K demonstrate the absence of sharp phonon linewidths, indicating that finite-size effects are relatively unimportant. Also, we note that systems with these dimensions provide reliable estimates of thermal conductivity using Green-Kubo approaches.

\begin{figure}
\begin{centering}
\includegraphics[scale=0.75]{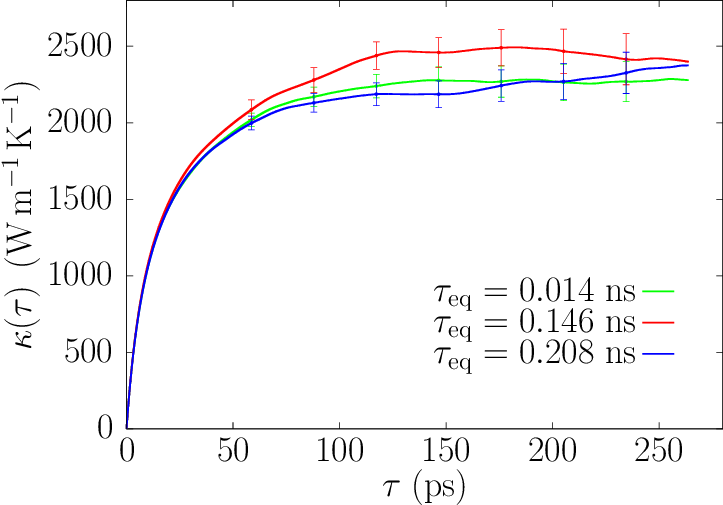} 
\caption{Thermal conductivity integral in Eq. \ref{GK} plotted as a function of the upper integration limit $\tau$ for different equilibration times $\tau_{eq}$.
}
\label{fig1}
\end{centering}
\end{figure}

In addition to requiring a large statistical ensemble, the calculation of $\kappa$ may depend somewhat on the initial equilibration period under the action of the thermostat. Each of the $640$ simulations in the ensemble was preceded by an independent, constant temperature equilibration period. 
In Fig. \ref{fig1}, the value of $\kappa(\tau)$ is shown as a function of the upper limit of integration $\tau$ for three different equilibration times. The error bars in Fig. \ref{fig1} represent the standard deviation determined over the entire ensemble. While there does appear to be some weak dependence on $\tau_{eq}$, for $\tau_{eq}=0.146$ns and $\tau_{eq}=0.208$ns, the results agree to within the statistical uncertainty indicated by the error bars. For calculation of response functions, it will be demonstrated that the results involve shorter time scales and are even less clearly dependent on $\tau_{eq}$. In the following, $\tau_{eq}=0.208$ns was used for all reported data.

The room-temperature thermal conductivity calculated using Eq. \ref{GK} in this work is $2370 \pm 150$ W/mK. Using the same optimized Tersoff potential, Fan reported a $\kappa$ value of $2700 \pm 80$ W/mK in Ref. \cite{Fan:2015tj}. While these results are comparable, one possible reason for the disagreement might be the high aspect ratio of our simulation supercell in comparison to the supercell structures used by Fan \cite{Fan:2015tj}.    The calculated $\kappa$ in this paper agrees well with  $\kappa \approx$ 2300 W/mK obtained using the same Tersoff potential with  non-equilibrium MD simulation (NEMD) by Xu \cite{Xu_2014}. Other results by Haskins in Ref.\cite{Haskins:2011aa} calculate values of $\kappa$ around 2600 W/mK using an Einstein relation and MD simulation. The $\kappa$ result given here is also comparable to experimental value $\kappa \approx 2500 \pm 1000$ W/mK at 350 K obtained from Raman spectroscopy of suspended CVD grown graphene \cite{Cai_2010}.

The main focus of this paper is the computation of thermal response-functions which were first described in  Ref \cite{Fernando_2020}. The basic idea is to start from the perspective that
an externally-input power density $H^{(ext)}(\bm{x}^{\prime},t^{\prime})$ will result in a subsequent heat-current density $J_{\mu}(\bm{r},t)$ according to,
\begin{equation} \label{def1}
J_{\mu}(\bm{r},t)= -{1 \over C_{V}} \int_{-\infty}^{t} dt^{\prime} \int_{\Omega} d^{3} r^{\prime}  K_{\mu \mu} (\bm{r}-\bm{r}^{\prime},t-t^{\prime}) {\partial H^{(ext)}(\bm{r}^{\prime},t^{\prime}) \over \partial r^{\prime}_{\mu}} \text{  ,}
\end{equation}
which assumes linear response, but crucially does not assume diffusive transport. Here, the subscript $\mu$ refers to a Cartesian component, and the expression in Eq. \ref{def1} assumes that the axes are chosen to reflect the crystal symmetry such that the input energy gradient results in a current along the same Cartesian direction $\mu$. If the Cartesian axes are not chosen in this way, then the response function tensor $\bm{K}$ will have non-zero off-diagonal components.  Finally, in Eq. \ref{def1}, $C_{V}$ is the system heat capacity and $\Omega$ is the system volume. In the case of a two-dimensional system like graphene, $\Omega$ includes the
graphite layer spacing, just as with the Green-Kubo calculations of $\kappa$ described earlier.

In a periodic system, it is possible to write Eq. \ref{def1} entirely in Fourier space. We additionally take the external heat source to be a pulse at $t^{\prime}=0$ represented by,
$H^{(ext)}_{\bm{q}}(t^{\prime}) = u^{(ext)}_{\bm{q}}(0) \delta(t^{\prime})$. Then, with the assumption that the system is in equilibrium before the heat pulse, we can
express the resulting current as,
\begin{equation} \label{def2}
J_{\bm{q}}(t) = -{i \over c_{V}} q K_{\bm{q}}(t) u^{(ext)}_{\bm{q}}(0)
\end{equation}
in which $c_{V}={C_{V} \over \Omega}$ is the volumetric specific heat capacity and $q$ is the magnitude of the vector $\bm{q} = q \hat{e}_{\mu}$.
In practical MD calculations using periodic-boundary conditions, the possible vectors $\bm{q}$ must correspond to reciprocal-lattice vectors for the simulation supercell.

Following the arguments in  Ref. \cite{Fernando_2020} and Ref. \cite{Schelling:2025aa}, we can determine the response function by computing the dissipation of thermal fluctuations within the equilibrium ensemble. The response function can be determined for a particular $\bm{q}$ vector from,
\begin{equation} \label{corrMD}
 K_{\bm{q}}(\tau) =c_{V} {\int_{0}^{\tau} dt \langle J_{\bm{q}}(t) J_{-\bm{q}}(0)\rangle  \over \langle  u_{\bm{q}}(0) u_{-\bm{q}}(0)\rangle}
 \end{equation}
 To compute the Fourier components $J_{\bm{q}}$, we require as a starting point a local definition for the heat current, which is also directly related to the
 Fourier components $u_{\bm{q}}$ of the thermal energy density. These details were presented previously in Ref. \cite{Schelling:2025aa}. Finally, the approach here ignores ``convective'' terms in the definitions, which are generally understood to be unimportant for heat transport in crystalline systems. To reflect on 
 the relevance of the convective term, we first note that, in terms of quantum creation and annihilation operators, the convective term is actually cubic in contrast to the ``virial'' part which is quadratic. The other relevant fact is that the convective part, which is the product of an energy density times a velocity vector, results in rapid oscillations at phonon frequencies, which are not directly connected to the slow relaxation of a thermal fluctuation. We have shown \cite{Schelling:2025aa}  that relaxation of a periodic perturbation with wave-vector $\bm{q}$ results from the virial part of heat conduction and depends on differences in phonon frequencies, and hence results in lower-frequency response. In short, the convective terms are lower order and result in rapid oscillations that are not directly related to the relaxation of a perturbation.
 
Response functions that presume the validity of Fourier's law and the heat-diffusion equation are easily established as Green's functions. Specifically, assuming
the validity of the heat-diffusion equation, it can be shown that $K_{\bm{q}}(\tau) = \kappa e^{-\alpha q^{2} \tau}$. Here $\kappa$ is the thermal conductivity and $\alpha={\kappa \over c_{V}}$ is the thermal diffusivity. This provides a basis to establish deviations from the heat-diffusion equation.

\section{Results: Thermal Response Functions}

In the following, the data was obtained from the same ensemble of $T=300$K simulations for graphene. As mentioned earlier, the choice of a system with dimensions $320 \times 32$ primitive cells was made so that transport could be studied for fairly small values $q=|\bm{q}|$ while keeping the computational cost moderate. Given the weak dependence on $\tau_{eq}$ found in Fig. \ref{fig1}, we used the data obtained from $\tau_{eq}=0.208$ ns which appears well converged. In the following, we will also show directly that the results are very
weakly dependent on $\tau_{eq}$.

\begin{figure}
\begin{centering}
\includegraphics[scale=0.75]{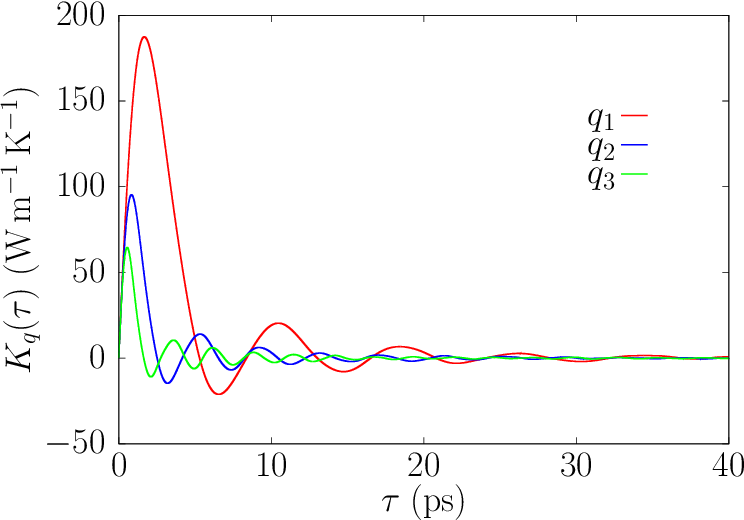} 
\caption{Response function $K_{\bm{q}}(\tau)$ plotted as a function of time for three different wave vectors with magnitudes $q_{1}={2 \pi \over L}$, $q_{2}={4 \pi \over L}$, and $q_{3}={6 \pi \over L}$. As described in the text, the length scale is $L=68.1$nm.
}
\label{fig2}
\end{centering}
\end{figure}

In Fig. \ref{fig2}, we show the computed $K_{\bm{q}}(\tau)$ at $T=300$K for three different values of the wave vector, specifically $q_{1}={2 \pi \over L}$, 
$q_{2}={4 \pi \over L}$, and $q_{3}={6 \pi \over L}$ with $L=68.1$nm as the long dimension of the supercell. These results demonstrate very clear oscillatory heat propagation. For example, with $q_{1}={2 \pi \over L}$, at least four oscillation periods are clearly discernible. For larger magnitudes $q_{2}$ and $q_{3}$, the oscillation period and decay time both decrease. We will return to these observations later on.

\begin{figure}
\begin{centering}
\includegraphics[scale=0.75]{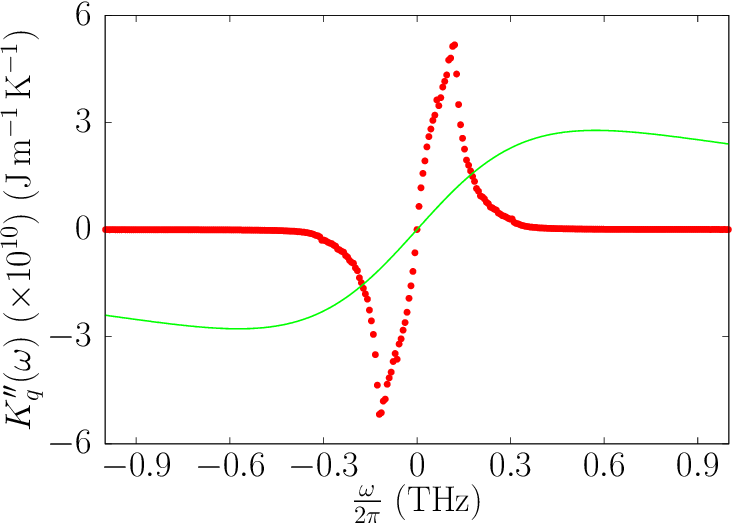} 
\caption{Imaginary part $K^{\prime \prime}_{\bm{q}}(\omega)$ of the Fourier transformed response function plotted as a function of frequency ${\omega \over 2 \pi}$ for $q_{1}={2 \pi \over L}$. The green curve shows the predicted curve based on Fourier's law using the classical heat capacity and computed thermal conductivity values $\kappa$.
}
\label{fig3}
\end{centering}
\end{figure}

\begin{figure}
\begin{centering}
\includegraphics[scale=0.75]{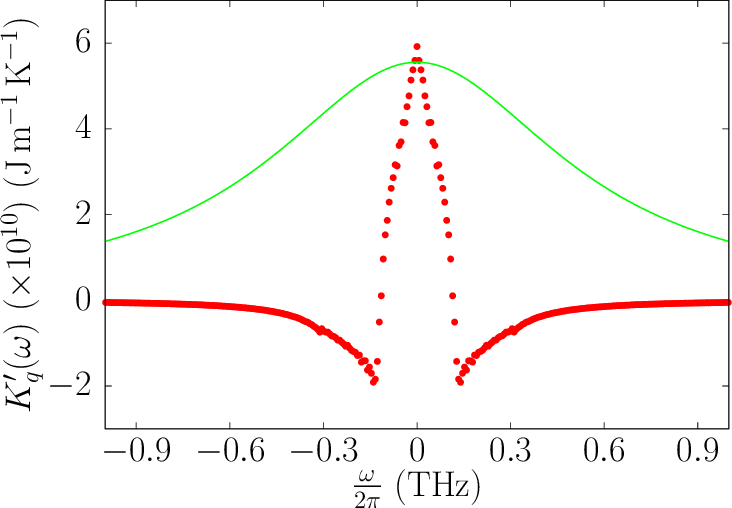} 
\caption{Real part $K^{\prime}_{\bm{q}}(\omega)$ of the Fourier transformed response function plotted as a function of frequency ${\omega \over 2 \pi}$ for $q_{1}={2 \pi \over L}$. The green curve shows the predicted curve based on Fourier's law using the classical heat capacity and computed thermal conductivity values $\kappa$.}
\label{fig4}
\end{centering}
\end{figure}

To identify spectral features connected to the normal modes, we next turn to the Fourier transform of the response function,
\begin{equation}
\tilde{K}_{\bm{q}}(\omega) = \int_{0}^{\tau_{m}} K_{\bm{q}}(\tau) e^{i \omega \tau}d\tau
\text{  .} 
\end{equation}
Here the upper limit of integration $\tau_{m}=0.262$ns is the maximum time for each member of the simulation ensemble. This time primarily limits the 
resolution of the Fourier transform to $3.4 \times 10^{-3}$THz. 
While $ K_{\bm{q}}(\tau) $ is a real function, the Fourier transform results in real and imaginary components $K^{\prime}_{q}(\omega)$ and $K^{\prime \prime}_{\bm{q}}(\omega)$ respectively
so that $\tilde{K}_{\bm{q}}(\omega) = K^{\prime}_{q}(\omega)+i K^{\prime \prime}_{\bm{q}}(\omega)$. In Fig. \ref{fig3}-\ref{fig4}, we show, respectively, the resulting data for $K_{\bm{q}}^{\prime \prime} (\omega)$ and $K^{\prime}_{q}(\omega)$. Along with the simulated data, the corresponding curves for Fourier's law and the heat diffusion equation using the simulated values of $\kappa$ and $\alpha$ are also presented. First, strong disagreement between the simulation results and the heat diffusion equation is immediately evident. In our previous studies of Lennard-Jones solids \cite{Fernando_2020} and hBN \cite{Schelling:2025aa}, we demonstrated that simulation results tend to converge towards the heat diffusion equation as $q$ decreases and/or temperature increases. Next, we note that for $K_{\bm{q}}^{\prime \prime} (\omega)$, the peaks occur near ${\omega \over 2 \pi} = \pm 0.12$ THz. This is in agreement with the oscillations in Fig. \ref{fig2} for $q_{1}$, which appear to have a period $\tau \approx 8$ps. Second, the width of the peaks in Fig. \ref{fig3} is $\sim 0.10$THz, which would correspond roughly to a lifetime of $\sim 5$ps. This quite short lifetime is consistent with the relatively strong damping of the oscillations in Fig. \ref{fig2}. Note that $K_{\bm{q}}^{\prime \prime} (\omega)$ is appropriate for this analysis since $K_{q}(\tau)$ is best described as a damped function $\sin(\omega \tau)$ curve. Further analysis in the context of the hyperbolic heat equation will follow later.

\begin{figure}
\begin{centering}
\includegraphics[scale=0.75]{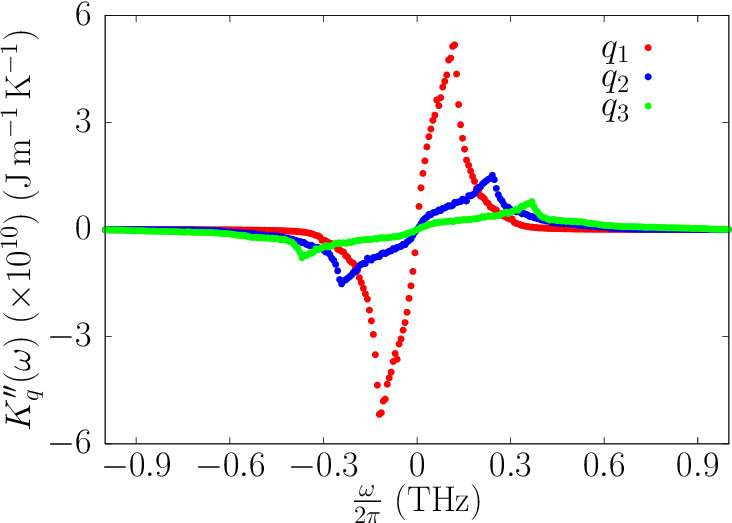} 
\caption{Imaginary component $K^{\prime \prime}_{\bm{q}}(\omega)$ plotted as a function of frequency for three different wave vectors with
magnitudes $q_{1}={2 \pi \over L}$, $q_{2}={4 \pi \over L}$, $q_{3}={6 \pi \over L}$ with $L=68.1$nm. 
}
\label{fig5}
\end{centering}
\end{figure}

\begin{figure}
\begin{centering}
\includegraphics[scale=0.75]{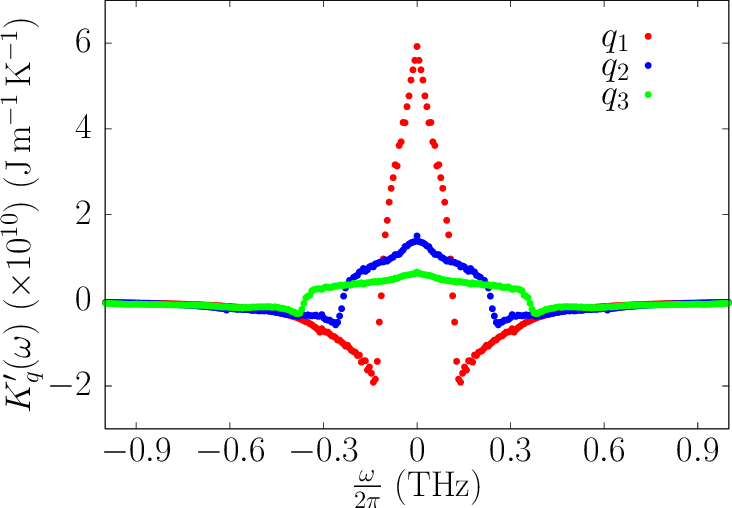} 
\caption{Real component $K^{\prime}_{\bm{q}}(\omega)$ plotted as a function of frequency for three different wave vectors with
magnitudes  $q_{1}={2 \pi \over L}$, $q_{2}={4 \pi \over L}$, $q_{3}={6 \pi \over L}$ with $L=68.1$nm. 
}
\label{fig6}
\end{centering}
\end{figure}

From our previous work \cite{Schelling:2025aa}, it is clear that larger magnitudes of the wave vector $q$ lead to a broader frequency range for the response and also sharper spectral features. Because the response is spread over a larger range of frequencies, sharp spectral features require less
resolution to observe, and are also not as limited by finite phonon lifetimes and anharmonic phonon scattering. In Fig. \ref{fig5} and \ref{fig6},
results for $K_{q}^{\prime \prime} (\omega)$ and $K_{q}^{\prime } (\omega)$  are shown to compare the spectra for each value $q_{1}={2 \pi \over L}$, $q_{2}={4 \pi \over L}$, and $q_{3} = {6 \pi \over L}$.  These general trends are as expected based on our understanding. Specifically, detailed interpretation of the results in terms of the phonon band structure will follow. In comparison to the results for hBN \cite{Schelling:2025aa}, there are fewer sharp spectral features here. However, the hBN results which exhibited very sharp, detailed spectra were obtained at temperatures as low as $T=100$K and for significantly shorter length scales.

\begin{figure}
\begin{centering}
\includegraphics[scale=0.75]{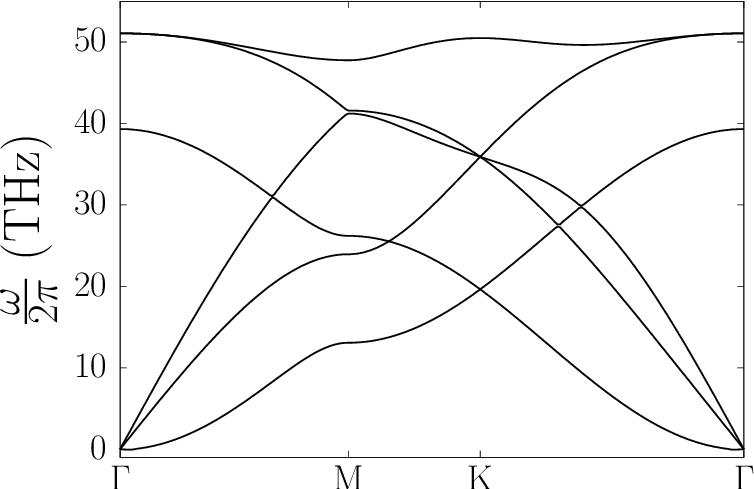} 
\caption{Computed phonon band structure for graphene. 
}
\label{fig7}
\end{centering}
\end{figure}

As in our previous study, the response in frequency space can be largely understood from the phonon band structure. In Fig. \ref{fig7}, the
computed band structure for the optimized Tersoff model \cite{Lindsay:2010vb} is shown. The data was generated using the phono3py software package
\cite{Togo_2023,Togo_2023.2} using calculated forces obtained from the empirical potential. In our previous articles \cite{Bohm:2022aa,Schelling:2025aa}, we have demonstrated that
for an excitation with wave vector $\bm{q}$, the response is characterized by ``beating'' with frequencies $\pm \left(\omega_{\bm{k} + {\bm{q} \over 2},s} - 
\omega_{-\bm{k} + {\bm{q} \over 2},s} \right)$ primarily within the same band $s$. This insight arises from the expression for $J_{\bm{q}}$ derived in Ref. \cite{Schelling:2025aa}
\begin{equation} \label{curr}
\bm{J}_{\bm{q}} = {1 \over 2\Omega} \sum_{\bm{k},s} \hbar \omega_{\bm{k}s} \bm{v}_{\bm{k}s} A_{\bm{k}+{\bm{q}\over 2},s}B_{-\bm{k}+{\bm{q}\over 2},s}
\text{  .}
\end{equation}
This expression was also derived by Hardy \cite{Hardy:1963td}. Here, the $A_{\bm{k}+{\bm{q}\over 2},s}$ and $B_{-\bm{k}+{\bm{q}\over 2},s}$
represent normal-mode coordinates that can be written, in a quantum-mechanical picture, in terms of phonon creation and annihilation operators. In a band with nearly linear dispersion, with $\bm{k}$ parallel to $\bm{q}$, we expect resonant behavior across a band with $\omega_{\bm{k} + {\bm{q} \over 2},s} - 
\omega_{-\bm{k} + {\bm{q} \over 2},s}\approx \pm \omega_{\bm{q},s}$. This allows a determination of where different phonon branches should contribute to $K^{\prime \prime}_{\bm{q}}(\omega)$. Hence the phonon spectra can be used to explain both the position of the maxima and the width of the $K^{\prime \prime}_{\bm{q}}(\omega)$ in frequency space. 

Given this insight, it is possible to identify the primary normal modes that contribute at different frequencies. For simplicity, the focus here is on interpretation of $K_{\bm{q}}^{\prime \prime}(\omega)$, although the same considerations apply to $K_{\bm{q}}^{\prime}(\omega)$. Analysis of the band structure in Fig. \ref{fig7} yields propagation speeds $v_{LA} \approx 20.3$ km s$^{-1}$ and $v_{TA} \approx 14.6$ km s$^{-1}$ for the longitudinal acoustic (LA) and transverse acoustic (TA) branches, respectively. For the parabolic ZA branch, $v_{ZA}$ approaches zero at the $\Gamma$ point, and has a maximum value $\sim 8$ km s$^{-1}$. In the case of LA and TA modes with linear dispersion, the maximum frequencies that contribute to the response function can be determined. For $q_{1}={2 \pi \over L}$, the prediction is for response at a maximum frequency $\sim 0.30$ THz due to the LA branch, which is quite close to a small peak visible at $0.31$ THz. For TA contributions, the maximum frequency is predicted to be $\sim 0.22$ THz. The large peak in the response that occurs at $\sim 0.12$ THz is therefore primarily attributable to the ZA branches which should contribute significantly in the range $0-0.16$THz. However, optical modes and also other acoustic modes propagating with $\bm{k}$ with a component perpendicular to $\bm{q}$ also contribute. These considerations work equally well for $q_{2}$ and $q_{3}$ data at shorter length scales.

\begin{figure}
\begin{centering}
\includegraphics[scale=0.75]{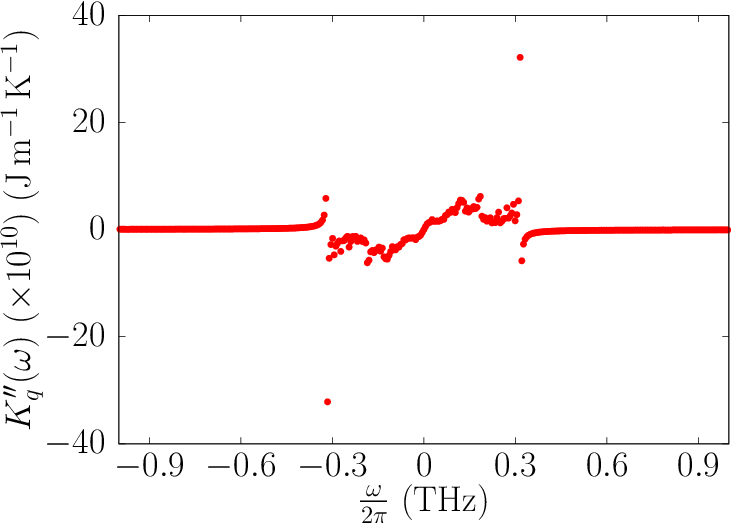} 
\caption{Response function $K_{\bm{q}}^{\prime \prime}(\omega)$ for $q_{1}={2 \pi \over L}$ with $L=68.1$nm computed at temperature $T=10$K.
}
\label{fig8}
\end{centering}
\end{figure}

\begin{figure}
\begin{centering}
\includegraphics[scale=0.75]{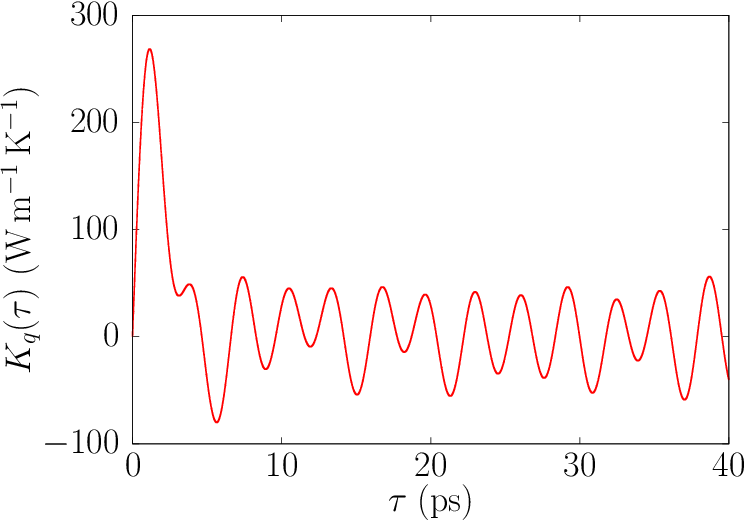} 
\caption{Time dependent response function 
$K_{\bm{q}}(\tau)$ for $q=q_{1}={2 \pi \over L}$ computed at temperature $T=10$K. The result shows persistent
oscillations primarily due to resonance in the LA phonon branch.
}
\label{fig9}
\end{centering}
\end{figure}

While the spectral width of the response functions is controlled primarily by the vector $\bm{q}$ and the phonon dispersions, the sharpness of resonant features depends on anharmonic scattering. We first present results for $K_{\bm{q}}^{\prime \prime}(\omega)$ computed at $10K$ for an ensemble of 640 runs each with 0.184ns of averaging time. For the wave same value $q_{1} = {2 \pi \over L}$, these results are shown in Fig. \ref{fig8}. While the overall width of the response function is essentially identical to the $T=300$K result in Fig. \ref{fig5}, more spectral detail is evident, including very sharp peaks corresponding to resonance in the LA phonon branch near $0.31$THz.  In Fig. \ref{fig9}, the time-dependent response function at $T=10$K is shown for $q_{1}={2 \pi \over L}$ demonstrating first rapid decay due to dispersion, followed by very long-lived, resonant oscillations in the LA band as a predominant feature. Taken together with the $T=300$K results, this demonstrates that sharp spectral features present at low temperatures become less evident with increasing temperature due to anharmonicity, but with limited effect on the overall width in frequency space of $\tilde{K}_{\bm{q}}(\omega)$.

\begin{figure}
\begin{centering}
\includegraphics[scale=0.75]{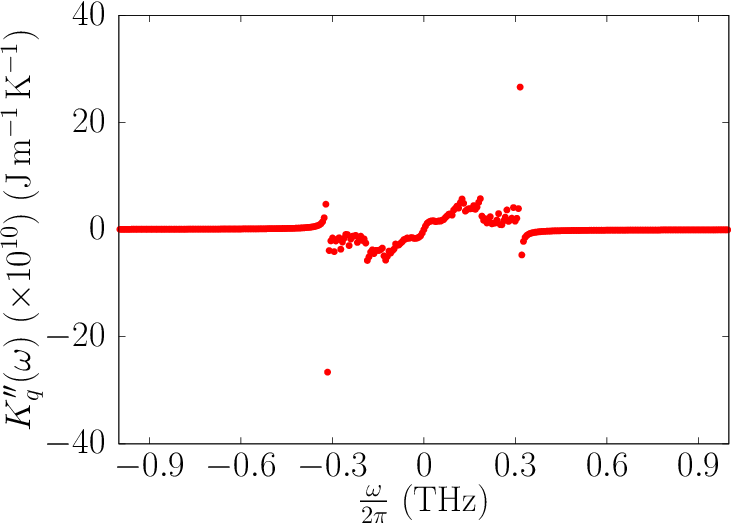} 
\caption{Response function $K_{\bm{q}}^{\prime \prime}(\omega)$ for $q_{2}={4 \pi \over 2L}$ with $2L=136.2$nm computed at temperature $T=10$K.
}
\label{fig10}
\end{centering}
\end{figure}

\begin{figure}
\begin{centering}
\includegraphics[scale=0.75]{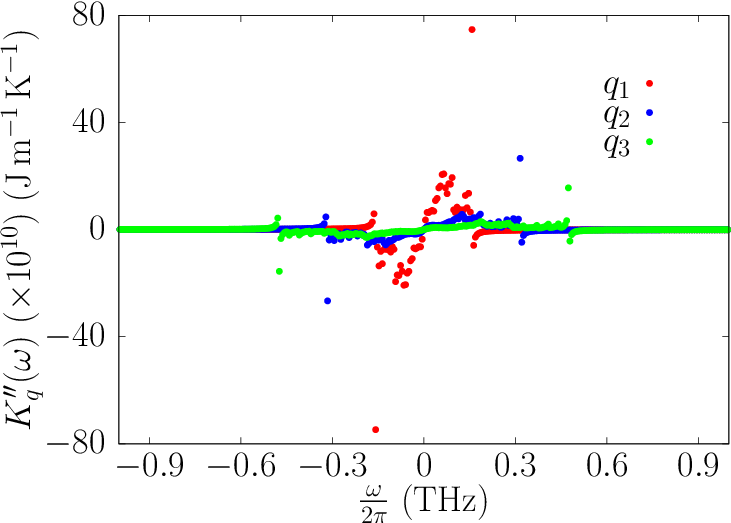} 
\caption{Response function $K_{\bm{q}}^{\prime \prime}(\omega)$ for $q_{1}= {2 \pi \over 2L}$, $q_{2}={4 \pi \over 2L}$ and $q_{3}={6 \pi \over 2L}$ with the system size $2L=136.2$nm computed at temperature $T=10$K. Even that the largest scale, the response function shows detailed spectra including a very sharp peak associated with the LA branch.
}
\label{fig11}
\end{centering}
\end{figure}

Another question is related to scaling of the response function with system size. Specifically, we investigated to what extent the response functions depend on system size $L$ for the same wave vector $\bm{q}$. To do this, we simulated at $T=10$K a system with twice the length in comparison to the system described above. Specifically, the system size was $2L=136.2$nm but with the same width in the transverse direction and $N=40,960$ atoms. Results were obtained using a slightly smaller ensemble of 420 independent members each with 0.184ns averaging time. Hence the total averaging time was 77.28ns. The results for $K^{\prime \prime}_{q}(\omega)$ for $q={4 \pi \over 2L }= {2\pi \over L}$ are shown in Fig. \ref{fig10}. These results are directly comparable to Fig. \ref{fig8}. The comparison is nearly exact indicating that scaling with system size is a relatively small effect at least for these rather large dimensions. The exception is the amplitude of the large LA peak near $0.31$THz, which is somewhat lower for the larger system in comparison to the smaller system size. Specifically the peak in the $L=68.1$nm results was $\sim 32.0 \times 10^{-10}$Jm$^{-1}$K$^{-1}$, while for $L=136.2$nm the value was $\sim 26.6\times 10^{-10}$Jm$^{-1}$K$^{-1}$. Statistical error was found to be $\sim \pm 1.2  \times 10^{-10}$Jm$^{-1}$K$^{-1}$. Hence, the difference in this one point appears to be real. There are several possibilities. First, we note that the frequency resolution was not adequate to resolve the peak, and hence any small shift in the frequency due to finite-size effects might affect the Fourier transform. However, we can nevertheless conclude that the qualitative results are not impacted by system size, and indeed the quantitative differences appear only weakly system-size independent. Finally, we show three different $q$-vectors for the system with $L=136.2$nm all computed at $T=10$K in Fig. \ref{fig11}. As with the other cases, the clear dependence of the spectral width on $\bm{q}$ is shown, with detailed spectral features evident even for $q_{1}= {2 \pi \over 2L}$. The location of the very sharp LA branch response scales depends linearly on $\bm{q}$ exactly as expected. This is further evidence of the dependence of the spectral width on both $\bm{q}$ and on details of phonon band structure. 

We next analyze some of the relevant phonon lifetimes at both $T=10$K and $T=300$K. 
To this end, we have computed velocity autocorrelation functions in reciprocal space. Specifically, for graphene with two atoms per unit cell, $m=1,2$, we compute  the velocity Fourier transform with the $N$ unit cells labeled by $l$,
\begin{equation}
\bm{v}_{m \bm{k}}(t) = {1 \over \sqrt{N}} \sum_{l=1}^{N} \bm{v}_{ml}(t) e^{i \bm{k} \cdot \bm{R}_{l}}
\end{equation}
in which $\bm{R}_{l}$ is a Bravais lattice vector. We compute the normal-mode spectrum for $\bm{k}$ from the time-averaged correlation function and summed over both species $m=1,2$,
\begin{equation} \label{zeq}
Z_{\bm{k}}(\omega) = \sum_{m=1,2} \int_{0}^{\tau_{max}} d \tau e^{i \omega \tau} \langle \bm{v}^{\ast}_{m \bm{k}}(\tau) \cdot
\bm{v}_{m \bm{k}}(0)  \rangle
\end{equation}
This quantity was computed from an ensemble average of $8$ independent runs with $192.5$ps of integration time for each ensemble member.

\begin{figure}
\begin{centering}
\includegraphics[scale=0.75]{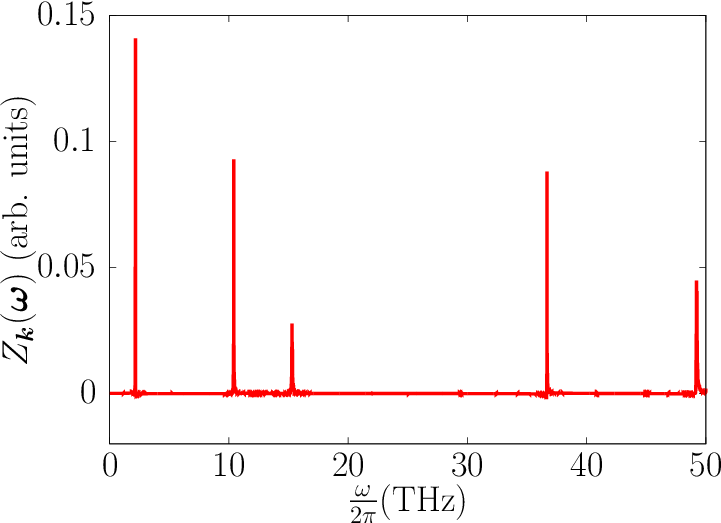} 
\caption{The quantity $Z_{\bm{k}}(\omega)$ determined from Eq. \ref{zeq} for graphene at $T=300$K. The wave vector $\bm{k}$ was chosen along the $\Gamma \rightarrow M$ direction, with magnitude $k={ 2\pi \over 3a} {5 \over 16}$.
}
\label{fig12}
\end{centering}
\end{figure}

\begin{figure}
\begin{centering}
\includegraphics[scale=0.75]{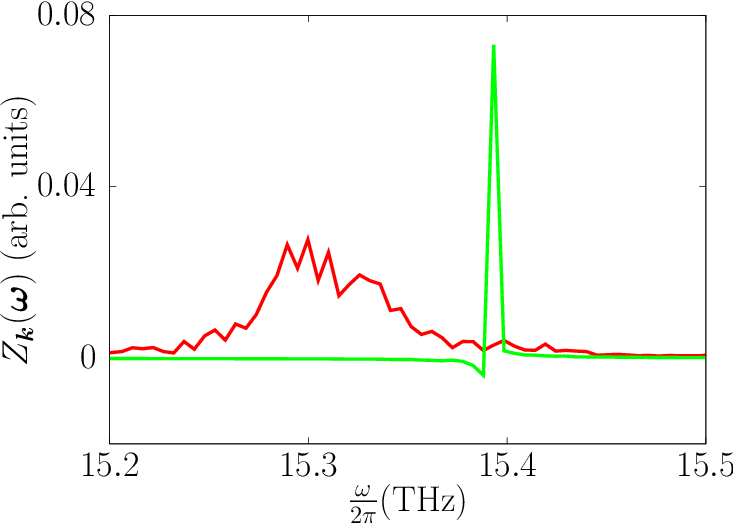} 
\caption{The quantity $Z_{\bm{k}}(\omega)$ determined from Eq.  \ref{zeq} for graphene at $T=300$K (red curve) and $T=10K$ (green curve). The wave vector $\bm{k}$ was chosen along the $\Gamma \rightarrow M$ direction, with magnitude $k={ 2\pi  \over 3a} {5 \over 16}$. The frequency range was chosen to highlight the peak associated with the LA phonon branch.
}
\label{fig13}
\end{centering}
\end{figure}

We first focus on the overall spectrum for $|\bm{k}|={2 \pi \over 3a}{5 \over 16}$ along $\Gamma \rightarrow M$ at $T=300$K shown in Fig. \ref{fig12}. The ZA, TA, LA, and optical peaks are clearly evident. The linewidths were analyzed and hence the lifetimes determined for these typical results. First, it is apparent that the lifetime of the LA branch is the smallest in comparison to the others. In Fig. \ref{fig13}, the LA peak at both $T=10$K and $T=300$K is shown for comparison. At $T=300$K, while the width of the LA peak is significant, it is still much less than the width of the response function $K_{\bm{q}}^{\prime \prime}(\omega)$ shown in Fig. \ref{fig5}. At $T=300$K, the estimated lifetime is $\tau_{LA}=2.7$ps. In contrast, the other acoustic branches have rather narrow linewidths with corresponding lifetimes $\tau_{ZA}=19.9$ps and $\tau_{TA} = 15.9$ps. These lifetimes are reasonably comparable to first-principles DFT results with quantum statistics reported in Ref. \cite{Bonini_2012}. This point is discussed below. For the LA peak at $T=10$K, the linewidth is beneath the resolution of the calculation, indicating that $\tau_{LA}$ is at least $\sim 200$ps. This agrees with the very sharp LA feature in Fig. \ref{fig8}.

In summary of these observations, the overall width of the response functions in frequency space is primarily controlled by the vector $\bm{q}$ and phonon dispersion. As $\bm{q}$ increases in magnitude, the width of $\tilde{K}_{\bm{q}}(\omega)$ predictably increases, as the data in Figs. \ref{fig5}-\ref{fig6} reveals.  However, at low temperatures, very sharp spectral features are observed. Specifically, we have shown that at $T=10$K, sharp resolved features, including in particular dramatic peaks corresponding to resonance in the LA branch, are observed. While these sharp peaks do not result in substantial  changes in the overall width of $\tilde{K}_{\bm{q}}(\omega)$, they indicate the presence of long-lived resonance only within the LA branch, rather than more broadly across the entire spectral range. Because resonances of this kind involve a specific phonon branch, other authors often call this response ``first sound'' rather than second sound, but here we do not draw that distinction. As temperature increases toward $T=300$K, these sharp spectral features broaden. At the point where sharp resonant features are lost, which is
apparently the case at $T=300$K for graphene, the oscillation lifetime is determined almost entirely by phonon dispersion and the width of the response function $\tilde{K}_{\bm{q}}(\omega)$.

Another question is the effect of classical statistics on the response functions. First, we note that the phonon lifetimes we obtained from the analysis at $T=300$K are in reasonable agreement with first-principles results from Ref. \cite{Bonini_2012}. For example, the shortest lifetime LA modes we estimate $\tau_{LA}=2.70$ps, where as in Ref.  \cite{Bonini_2012}, the LA branch has lifetimes below $\sim 5$ps. This indicates that phonon linewidths, while different from first-principles results, are not the principal source of error. More significant is the fact that quantum statistics should essentially eliminate contributions from higher frequencies. For $T=300$K, we can estimate a frequency $f={k_{B}T \over h} =6.25$THz where quantum statistics should play a crucial role. Given the very large bandwidth for graphene in Fig. \ref{fig7}, most of the normal modes in graphene are not described by the classical heat capacity. 

To estimate the response function curves with quantum statistics, we note that the theory in Ref. \cite{Schelling:2025aa} showed that, apart from
correlated scattering events, the response functions are proportional to the mode-dependent heat capacity multiplied by the square of the phonon velocity. Therefore, for each frequency $\omega$, we computed a scale factor $W(\omega)$,
\begin{equation} \label{weight}
W(\omega) = \frac{\sum_{\bm{k} s} C_{V}(\omega_{\bm{k}s},T) v_{\bm{k}s,\mu}^{2} 
 \delta_{\omega,\omega_{\bm{k}+{\bm{q} \over 2 }s}-\omega_{\bm{k}-{\bm{q} \over 2 }s}}} 
 {\sum_{\bm{k} s} k_{B}  v_{\bm{k}s,\mu}^{2} \delta_{\omega,\omega_{\bm{k}+{\bm{q} \over 2 }s}-\omega_{\bm{k}-{\bm{q} \over 2 }s}} }
\end{equation}
in which $ C_{V}(\omega_{\bm{k}s},T) v_{\bm{k}s,\mu}^{2} $ is the product of the quantum heat capacity and squared velocity component $\mu$ for a normal mode with wave-vector $\bm{k}$ in branch $s$. The Kroenecker delta functions are used to select the frequency $\omega$ based on the resonant frequency 
$\omega_{\bm{k}+{\bm{q} \over 2 }s}-\omega_{\bm{k}-{\bm{q} \over 2 }s}$. To correct for quantum effects in this approximate way,  the response function $\tilde{K}_{\bm{q}}(\omega)$ obtained from classical MD simulation was multiplied by the weighting function $W(\omega)$. This is approximate in part because 
it is not possible to selectively filter out the contributions of different parts of the phonon spectra. Rather, this provides an approximate overall scaling factor at each frequency $\omega$.

\begin{figure}
\begin{centering}
\includegraphics[scale=0.75]{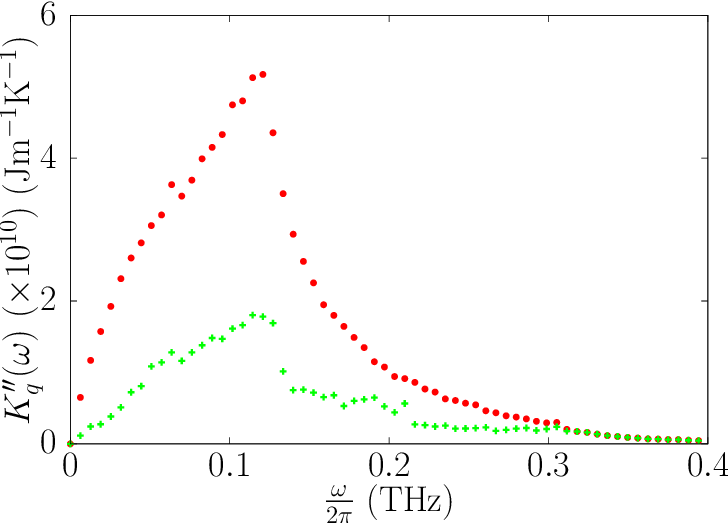} 
\caption{Response function $K_{\bm{q}}^{\prime \prime}(\omega)$ for $q=q_{1}={2 \pi \over L}$ at $T=300$K (red circles) compared
alongside the curve approximately corrected for quantum statistics (green crosses) using  the weight factors from Eq. \ref{weight} as explained in the text.
}
\label{fig14}
\end{centering}
\end{figure}

The comparison of the classical and quantum-corrected curves for $K_{\bm{q}}^{\prime \prime}(\omega)$ are shown just for positive frequencies in Fig. \ref{fig14}. First, as expected, the
quantum statistics decreases the response function magnitude. The exception is the tail above about $0.30$THz which is entirely due to the LA phonon branch and hence is low-frequency and reasonably well described classically. The overall decrease in magnitude is expected due to the fact that quantum mechanics ``freezes out'' much of the vibrational spectra in graphene at $T=300$K. However, qualitatively, the width of the response function is very similar in both cases, suggesting that predicted lifetimes for oscillatory transport are not dramatically impacted by classical statistics. Specifically, in both the classical and quantum-corrected cases, the width of the response function is $\sim 0.10-0.12$THz which suggests coherent transport with lifetimes $\sim 1.3-1.6$ ps. The question of lifetimes is addressed in the next section.

\begin{figure}
\begin{centering} 
\includegraphics[scale=0.75]{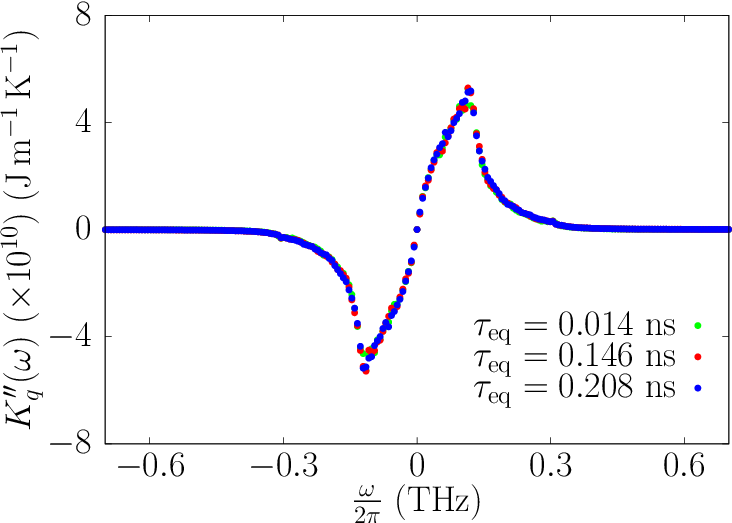} 
\caption{Response function $K_{\bm{q}}^{\prime \prime}(\omega)$ for $q=q_{1}={2 \pi \over L}$ obtained from data sets corresponding
to the three different times $\tau_{eq}$ for the equilibration step. This plot demonstrates that results are not sensitive to $\tau_{eq}$.
}
\label{fig15}
\end{centering}
\end{figure}

Finally, to establish that the results above are independent of $\tau_{eq}$, the simulated equilibration time, we show in Fig. \ref{fig15} results
for $K_{\bm{q}}^{\prime \prime}(\omega)$ for $q_{1}={2 \pi \over L}$ for each of the three values of $\tau_{eq}$. As the figure demonstrates,
the results are essentially independent of $\tau_{eq}$. This justifies the use of $\tau_{eq}=0.208$ns results for the analysis above.

\section{Analysis}

\begin{figure}
\begin{centering}
\includegraphics[scale=0.75]{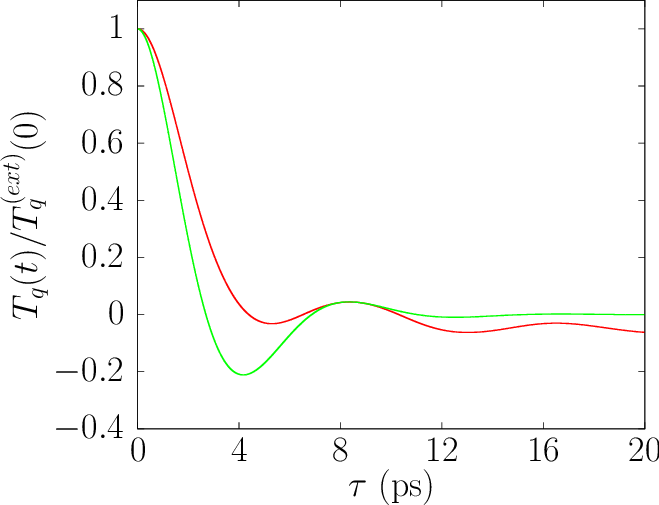} 
\caption{Computed component $T_{\bm{q}}(t)/T^{(ext)}_{\bm{q}}(0)$ obtained using Eq. \ref{tevolve1}  (red curve) with fit obtained using the solution
of the hyperbolic heat equation (green curve) given by Eq. \ref{hheqsoln}.  
}
\label{fig16}
\end{centering}
\end{figure}

We next apply the expression in Eq. \ref{hheq2} to the response-function data to obtain values for $\tau_{ss}$ and $v_{ss}$ that best fit the simulation data.
Moreover, here we will develop the equation for evolution of the temperature field that can be directly compared to experiments using TTG techniques.
Assuming that Eq. \ref{hheq2} is in the underdamped regime where $\tau_{ss} < {1 \over 2qv_{ss}}$, the solutions for $T_{\bm{q}}(t)$ given 
an initial pulse $T^{(ext)}_{\bm{q}}(0)$ at a specific wave vector $\bm{q}$, are,
\begin{equation} \label{hheqsoln}
T_{\bm{q}}(t) = T^{(ext)}_{\bm{q}}(0)  { e^{-{t \over 2\tau_{ss}}}\cos {\left(\beta t - \phi \right)} \over \cos \phi} \text{   ,}
\end{equation}
in which the oscillation frequency $\beta = \sqrt{\omega_{q}^{2} -{1 \over 4 \tau_{ss}^{2}}}$ and $\omega_{q}=v_{ss}q$.  The angle $\phi$ is chosen so that ${dT_{\bm{q}} \over dt}=0$ at $t=0$ which corresponds exactly to the MD simulation results. Then for $\phi$ we have the expression,
\begin{equation}
\tan \phi = {1 \over 2 \beta \tau_{ss}}  \text{   .}
\end{equation}
 From the computed response functions $K_{\bm{q}}(\tau)$, the Fourier component $T_{\bm{q}}(t)$ due to 
an initial perturbation $T^{(ext)}_{\bm{q}}(0)$ at $t=0$ evolves according to,
\begin{equation} \label{tevolve1}
T_{\bm{q}}(t) = T^{(ext)}_{\bm{q}}(0)\left(1-
{q^{2} \over c_{V}}\int_{0}^{t} K_{\bm{q}}(\tau) d\tau
\right)
\end{equation}
Using the results functions $K_{\bm{q}}(\tau)$ plotted in Fig. \ref{fig2}, the quantity $1-
{q^{2} \over c_{V}}\int_{0}^{t} K_{\bm{q}}(\tau) d\tau$ can be computed as a function of the upper limit of integration, 
resulting in a prediction for ${T_{\bm{q}}(t) \over T^{(ext)}_{\bm{q}}(0)}$, showing the oscillations and the decay of the input heat
pulse. This quantity is shown for $q_{1}={2 \pi \over L}$ in Fig. \ref{fig16}. The curve is not very accurately described by Eq. \ref{hheqsoln}. Instead of attempting to fit  Eq. \ref{hheqsoln} directly to the simulation, we instead set the parameter $\beta$ to exactly give the correct time for one oscillation period. Then, the value of $\tau_{ss}$ is adjusted to fit exactly amplitude of the local maxima near $\sim 8$ps. While $\tau_{ss}$ is adjusted, the value of $v_{ss}$ is simultaneously varied to keep $\beta$ constant. 
This ``fit'' is also shown in Fig. \ref{fig16}.  This procedure was repeated for $q_{2}$ and $q_{3}$, resulting in values for $\tau_{ss}$ and $v_{ss}$ to determine trends which can be compared to experiment. These values are shown in Table 1. While this procedure is approximate, and the curve itself only qualitatively represents the simulation results, it serves to obtain values for $v_{ss}$ and $\tau_{ss}$ which can be compared to BTE methods. As we will see, BTE methods applied at this scale result in dramatically different values for $\tau_{ss}$ in particular, a fact which does not result from the specific fitting procedure used here. Also, we note that better phenomenological models exist. These will be discussed in the last section of the paper. However, we contend that the fundamental result that $\tau_{ss}$ is much smaller than BTE predictions, as we will see, is not a function of using a very simple phenomenological model. 

The results in Table 1 show, perhaps not surprisingly, that $v_{ss}$ depends only weakly on $q$, indicating approximately linear dispersion. This is despite the fact that, as has been shown, the 
oscillatory pulse is comprised of the effect of many phonon branches, many not characterized by linear dispersion. The lifetime $\tau_{ss}$ does show significant dependence on $q$. As demonstrated below, the strong dependence of $\tau_{ss}$ on $q$ is seen in experiment, but is not predicted by
the BTE theory of second sound.

The values in Table 1 can be discussed in the context of observed trends in the experimental TTG results for graphite \cite{Ding_2022}. In that work, grating periods ranged from $4-16 \mu$m which is substantially larger than the $68.1$nm scale discussed here. Nevertheless, the experimental  second sound speeds $v_{ss}$ were observed in the range between $3-4$ km s$^{-1}$, which is somewhat smaller than our values in Table 1. The trend in Ref. \cite{Ding_2022} is for $v_{ss}$ to increase as the grating period decreases, especially for values below $\sim 3 \mu$m. However, experimental values for $v_{ss}$ vary by less than a factor of $2$ for grating periods from $3-15 \mu$m. However, the trend of increasing $v_{ss}$ with decreasing grating period is broadly consistent with the larger values for $v_{ss}$ for the length scales simulated here.

In addition, the lifetime $\tau_{ss}$ is reported for experimental results with graphite samples in Ref. \cite{Ding_2022}. 
 Specifically, Ref. \cite{Ding_2022} reports propagation lengths $\lambda_{ss}$, defined by $\lambda_{ss} =  2v_{ss} \tau_{ss}$, which appear to approach zero as the grating period approaches zero, and saturate at large grating periods to $\lambda_{ss}\sim 2 \mu$m. Hence, for grating periods below $\sim 4 \mu$m, experiment shows that $\tau_{ss}$ is strongly dependent on grating period. This is qualitatively consistent with the simulation data in Table 1, although admittedly simulation results were obtained for much smaller length scales.  For the smallest grating period $2 \mu$m used in the experiment,
 the value for $\tau_{ss}$ can be estimated from the data in Ref. \cite{Ding_2022} to be $\sim 20$ps. This $\tau_{ss}$ is larger than those in Table 1, but the trend is reasonable given the larger length scale in the experiment.  Hence, while the experiments were for a different material, graphite rather than graphene, and the length scales probed in the experiment were substantially longer than in our calculations, the experimentally-observed trends of $\tau_{ss}$ with grating period are at least qualitatively consistent with the theoretical understanding developed here.

\begin{table} 
\begin{center}
\caption{Resulting parameters $v_{ss}$ and $\tau_{ss}$ obtained from fitting simulation results to Eq. \ref{hheqsoln} as described in the text. The simulation data
and fit curve for $q_{1}$ are both shown in Fig. \ref{fig16}. Also included is the coherent propagation length $\lambda_{ss}=2v_{ss}\tau_{ss}$. The length scale $L=68.1$nm
for the MD simulations.
}
\vspace{0.5em}
\begin{tabular} {|c|c|c|c|}
\hline
$|\bm{q}|$ & $v_{ss}$ (km s$^{-1}$) & $\tau_{ss}$ (ps) & $\lambda_{ss}=2v_{ss}\tau_{ss}$ ($\mu$m) \\\hline 
$q_{1}={2 \pi \over L}$   & 9.12 & 1.34 & 0.024 \\\hline
$q_{2}={4 \pi \over L}$   &  8.69 & 0.96 & 0.016 \\\hline
$q_{3}={6 \pi \over L}$   & 8.68 &   0.68  & 0.011  \\\hline
\end{tabular}
\end{center}
\end{table}

Many studies of second sound have been based on solutions to the BTE. Early work by Enz \cite{ENZ1968114} classified second sound as either
``drifting'' or ``driftless''. Later, Hardy \cite{Hardy:1970aa} used the phonon BTE and showed that so-called drifting second sound requires normal scattering to dominate over Umklapp scattering, so that rapidly a drifting distribution is established. This condition is generally understood to be realized
in two-dimensional materials like graphene and is closely connected to both phonon hydrodynamics and second sound \cite{Cepellotti_2015,Lee_2015}. 
In Ref. \cite{Cepellotti_2015}, the BTE was solved in the Callaway approximation to the single-mode relaxation time approximation,
\begin{equation} 
\sum_{\nu_{2}} \Omega_{\nu_{1},\nu_{2}} n_{\nu_{2}} = -{n_{\nu}-n_{\nu}^{drift} \over \tau^{N}_{\nu_{1}}}
- {n_{\nu}-\bar{n}_{\nu} \over \tau^{R}_{\nu_{1}}} \text{    ,}
\end{equation}
in which $\nu_{1}=(\bm{k}_{1}, s_{1})$ and $\nu_{2}=(\bm{k}_{2}, s_{2})$, and $\Omega_{\nu_{1},\nu_{2}}$ is the phonon scattering matrix. The distribution $\bar{n}_{\nu}(T_{0})$ is the usual Bose-Einstein function
applied to each mode $\nu$. The drifted distribution is given by,
\begin{equation} 
n_{\nu}^{drift} = \left[e^{\beta (\hbar \omega_{\bm{k},s}- \hbar \bm{q} \cdot \bm{V})} -1 \right]^{-1}
\end{equation}
in which $\bm{q} \cdot \bm{V}$ represents a shift in the distribution which results from dominance of the momentum-conserving normal scattering. 
Using these approximations and also the continuity equation, the second-sound equation was derived in Ref. \cite{Cepellotti_2015,Lee_2015}, with expressions for $v_{ss}$ and $\tau_{ss}$. Specifically, for an isotropic medium, these were shown to be,
\begin{equation} \label{vss}
v_{ss}^{2} = {{1 \over 2} \sum_{\nu} C_{\nu} \bm{v}_{\nu} \cdot \bm{v}_{\nu} 
\over \sum_{\nu} C_{\nu}}
\end{equation}
\begin{equation} \label{tauss}
{1 \over \tau_{ss}} = {\sum_{\nu} { C_{\nu} q^{i}  v_{\nu}^{i} \over \omega_{\nu} \tau_{\nu}^{R}} 
\over \sum_{\nu} { C_{\nu} q^{i}  v_{\nu}^{i} \over \omega_{\nu} }} \text{  ,}
\end{equation}
in which $v_{\nu}^{i}$ and $q^{i}$ represent components of vectors Cartesian direction indicated by the superscript. It should be noted that here $q^{i}$ represents
a component of the ``drifted'' distribution in momentum space, and does not specifically refer a Fourier component of any excitation. Hence,
despite the presence of $q^{i}$ in Eq. \ref{tauss}, in fact the value of $\tau_{ss}$ is in fact independent of the specific value of $q^{i}$.  The scattering times $\tau_{\nu}^{R}$ are
obtained from the imaginary part of the self-energy function specifically related to Umklapp scattering process.
In Ref.  \cite{Cepellotti_2015}, this approach resulted in the predictions for graphene $\tau_{ss} \sim 100-200$ps and propagation lengths $\lambda_{ss}=2v_{ss}\tau_{ss} \sim 2\mu$m at $T=300$K \cite{Cepellotti_2015}. Note that the definition of $\lambda_{ss}$ in Ref.  \cite{Cepellotti_2015} differs by a factor of $2$ from the
one used here, which was chosen to agree with the definition in Ref. \cite{Ding_2022}.  It
is clear from Eq. \ref{tauss} that $\tau_{ss}$ depends on the resistive, Umklapp scattering rate $\tau_{\nu}^{R}$, weighted over all
of the phonon modes $\nu=(\bm{k}, s)$. In this picture then, it is only anharmonic scattering which causes decay of the drifting phonon distribution and the eventual dissipation of second sound.

\begin{figure}
\begin{centering}
\includegraphics[scale=0.75]{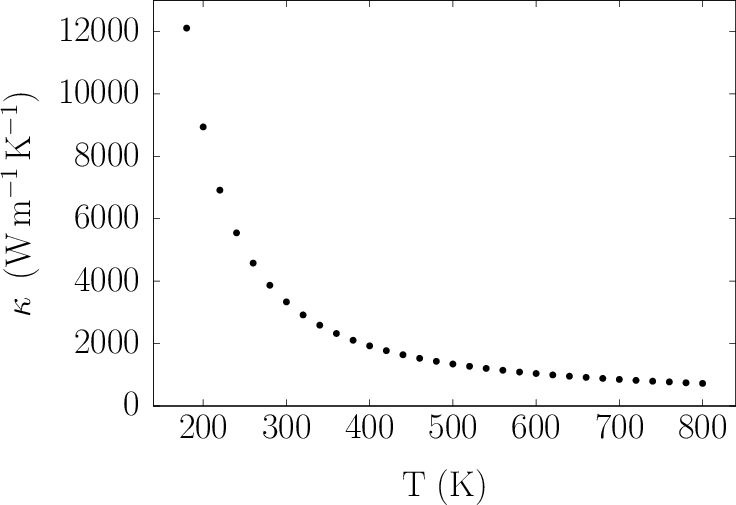} 
\caption{Graphene thermal conductivity $\kappa$ vs. temperature $T$ obtained from solutions of the LBTE.
}
\label{fig17}
\end{centering}
\end{figure}

Here we reproduce these results with our own calculations. We used the phono3py code \cite{Togo_2023,Togo_2023.2} along with VASP for DFT calculations\cite{Kresse_1994,KRESSE199615,Kresse:1999aa,PhysRevB.54.11169}. The system used was an $8 \times 8$ graphene supercell with $128$ atoms. The mesh in reciprocal space was a $3 \times 3 \times 1$ Monkhorst-Pack type mesh. The C-C bond length used in the DFT calculations was $1.425 \AA$. Gaussian smearing was used for the Brillouin zone integration with a sigma value of $\sigma = 0.05$ eV.
Thermal conductivity as a function of temperature $T$ was obtained by solving the full LBTE using a $100 \times 100 \times 1$ reciprocal-space mesh with the tetrahedron method for Brillouin-zone integration. The results
of these calculations are shown in Fig. \ref{fig17}. At room temperature, we obtained a value $\kappa=3,348$ Wm/K, which can be compared to the results of other authors. Specifically, also using VASP and the same reciprocal-space grid, Ref.  \cite{Han:2024} reports a room-temperature result $\kappa \approx 2,870$ W/mK at 300K. However, this result included isotopic scattering which likely accounts for their lower $\kappa$ value.  In Ref.\cite{Cepellotti_2015}, the LBTE was solve with a $128 \times 128 \times 1$ mesh and Gaussian smearing of 10 cm$^{-1}$. This work reports $\kappa \approx  4000 $W/mK near 300K which is comparable although slightly higher than our result.  

Having established the general validity of the DFT results, we next computed  $v_{ss}$ and $\tau_{ss}$  using the expressions in Eqs. \ref{vss}-\ref{tauss}. Our calculated values of $v_{ss}$ and $\tau_{ss}$ are 6.6 km/s and 50 ps resulting in a propagation length $\lambda_{ss}=2 v_{ss}\tau_{ss}=0.66 \mu$m. For comparison, Ref.\cite{Cepellotti_2015} reports a comparable value for $v_{ss} \sim 6.5$ km/s, but the somewhat larger value  $\tau_{ss} \sim 150$ ps. This leads to the propagation length $\lambda_{ss} \sim 2 \mu$m.  While these results are not in exact agreement, for reasons that are not immediately clear, they both result in propagation lengths $\lambda_{ss}$ at micron scales near room temperature.

From the analysis above, appears the BTE theory of second sound is in rather stark contrast with experiment and MD simulation results here for graphene and previously for hBN \cite{Schelling:2025aa}. This result does not imply that the BTE theory is incorrect, but rather that at short length scales like those probed by our simulation, phonon dispersion plays an important role even at a relatively high temperatures. The role of phonon dispersion is not captured by the BTE theory described above, and hence, while providing accurate predictions at longer length scales, may not capture important physics at shorter length scales. For lower temperatures, our results at $T=10$K also point to the potential for very sharp spectral features which would not be captured by the BTE theory. Sharp features of this kind have been called ``first sound'' by other authors rather than second sound.

At short length scales, the above results indicate that the BTE theory above results in substantially larger values for $\tau_{ss}$ in comparison to MD results. As noted elsewhere, TTG experiments on 
graphite demonstrate that $\tau_{ss}$ is strongly dependent on the grating period \cite{Ding_2022} which is qualitatively consistent with our interpretation that phonon dispersion and resulting loss of phase coherence with time is an important factor at least for shorter TTG periods and lower temperatures.  As length scales and/or temperature increases,
$\tau_{ss}$ should eventually be relatively insensitive to the TTG period. This is exactly what is seen in experiment.  For example, the experimental results for graphite \cite{Ding_2022} show that $\tau_{ss}$ saturates to a constant value at grating periods beyond $\sim 10 \mu$m. 

Other kinetic theories based on the Guyer-Krumhansl (GK) equation \cite{Guyer:1966aa,Sendra:2021aa,Sendra:2022aa} have been used to provide a description of the physics of second sound. While the above theory uses first-principles calculations, the reliance on the damped oscillator equation Eq. \ref{hheq2} for determination of $v_{ss}$ and $\tau_{ss}$ neglects viscosity and non-local effects that have been found to be important in interpreting experiments. For example, in Ref. \cite{Sendra:2021aa} the authors derived the GK equation based on a hypothesized deviation from equilibrium resulting in the evolution equation for the heat flux $\bm{J}(\bm{r},t)$,
\begin{equation}
\bm{J} = - \tau {\partial \bm{J} \over \partial t} - \kappa \bm{\nabla} T + l^{2} 
\left[ \nabla^{2} \vec{J} + \alpha^{\prime} \bm{\nabla}\left(\bm{\nabla} \cdot \bm{J} \right)\right]
\end{equation}
in which $\tau$ is the relaxation time, $\kappa$ is the bulk thermal conductivity, $l$ is a nonlocal length, and $\alpha^{\prime}$ is a nonlocal coefficient. The physical parameters of the GK theory were then subsequently computed using first-principles DFT methods. Importantly, the nonlocal length $l$ and relaxation time $\tau$ are described in terms of phonon velocities and relaxation times. For the present case, namely excitation by a sinusoidal source with a single vector $\bm{q}$, and using the continuity equation, the above equation can be rearranged and written in Fourier space,
\begin{equation}
{d^{2} T_{\bm{q}} \over dt^{2}} + {1 \over \tau}
\left [ 1+l^{2}q^{2}\left(1+ \alpha^{\prime}\right) \right]
{dT_{\bm{q} }\over dt} 
+ {\kappa q^{2} \over C_{v} \tau} T_{\bm{q}} = 0 \text{.}
\end{equation}
Hence for a TTG experiment, the GK theory is equivalent to the hyperbolic heat equation but perhaps with a different physical interpretation of the relaxation time.
Comparison with the hyperbolic heat equation shows that the lifetime of second sound is $\tau_{ss} = {\tau \over 1+l^{2}q^{2}(1+\alpha^{\prime})}$. Hence, for sinusoidal excitations, this theory does result in a lifetime $\tau_{ss}$ that depends on TTG period via $\bm{q}$.  In previous works, inclusion of the nonlocal term has been demonstrated to account for size effects \cite{Guo:2018aa}.   Another interesting study using the phenomenological Dual Phase Lag (DPL) model was able to show interplay between memory and non-local effects in computing the second-sound dispersion relation \cite{GANDOLFI2019118553}. However, the physics associated with this $\bm{q}$-dependence does not account for the phonon dispersion effects found in this paper as we discuss next.

The question then is whether we can definitively associate the bandwidth of the response functions, and then by extension the observed lifetime $\tau_{ss}$, on phonon dispersion effects rather than on the nonlocal physics in the GK theory. We stress again that we have unambiguously demonstrated the location of features in the response spectra associated with various phonon bands. This is perhaps most dramatically seen for the highest-frequency response due the the LA branch which yields a very sharp peak in the $T=10$K simulations. Response occurs due to resonances across the entire band structure at frequencies $\omega \sim \bm{v}_{\bm{k}s} \cdot \bm{q}$. This point was made in our recent work on hBN \cite{Schelling:2025aa}, and had been previously shown by Sham \cite{Sham:1967aa,Sham:1967ab}. Sharp resonant features are not predicted by the GK theory. Hence, while the nonlocal effects identified in the GK model are important, for shorter length scales we find that phonon dispersion effects are dominant. By contrast, at longer length scales, the GK theory predicts the most important physics and the effects identified in our simulations become less important.  Owing to these considerations, the simulations reported here at relatively short length scales are not able to identify nonlocal effects predicted by the GK model, but could be the focus of subsequent investigations.

At least one other approach using Green's functions established for the BTE  appears to include effects related phonon-dispersion has been reported  \cite{Chiloyan:2021uv}. In the reported experimental study of graphite in Ref. \cite{Ding_2022}, it was noted that the approach based on Eqs. \ref{vss}-\ref{tauss} above was unable to capture the dependence of $\tau_{ss}$ and $v_{ss}$ on the TTG period determined by $\bm{q}$. This observation is in agreement with our discussion and results of the MD simulations. Hence, instead of this approach, the BTE  first-principles based theory from Ref. \cite{Chiloyan:2021uv} was used, resulting in excellent agreement with the observed second sound. The connection between the theory in Ref.  \cite{Chiloyan:2021uv}  and the theory developed in our earlier work \cite{Schelling:2025aa} has not been entirely established. However, it should be noted that the theory in Ref. \cite{Chiloyan:2021uv} does predict spectral responses peaked at frequencies $\omega \sim \bm{q}\cdot \bm{v}_{\bm{k}s}$, and therefore we expect will generate spectra quite similar to our results in Figs. \ref{fig5}-\ref{fig6}. 

\section{Time-dependent sources}
In this section we elucidate oscillatory heat transport generated by time-dependent sources. We note that TTG studies generally involve a single pulse with short time duration occurring at a 
single wave vector $\bm{q}$ to describe the grating period. However, at least in principle, these excitations could be made to be time-dependent. In our original work \cite{Fernando_2020} we briefly discussed the response to time-dependent sources. Here the physics is elucidated in more detail, including predictions of where oscillatory transport might be either enhanced
or suppressed by changing the frequency $\omega$ of the excitation source for a specific $\bm{q}$. The main physical consideration here is the observation that the heat-current lags the excitation which generated it, and as a result, the transport is always out of phase with the excitation. This is in direct contrast to Fourier's law, where the current response is instantaneous.

Here we consider an excitation source given by,
\begin{equation}
H^{(ext)} (\bm{r},t)= {1 \over 4} H_{0}  \left(e^{i \bm{q} \cdot \bm{r}} +e^{-i \bm{q} \cdot \bm{r}}   \right) \left(e^{i \omega t} + e^{-i \omega t} \right)
=H_{0} \cos(\bm{q} \cdot \bm{r}) \cos(\omega t)
\end{equation}
Then following Ref. \cite{Fernando_2020}, the time-dependent temperature field, including the effect of the heating source, is given by,
\begin{equation}
T(\bm{r},t) = T_{0} + {H^{(ext)}_{0} \over c_{V} \omega}
\left[
\left(1-{q^{2} \over c_{V}} K_{q}^{\prime}(\omega) \right)
\sin\omega t   +{q^{2} \over c_{V}}K_{q}^{\prime \prime}(\omega) \cos\omega t 
\right]\cos(\bm{q} \cdot \bm{r})
\end{equation}
It is also useful to write this expression in a different form that shows the phase difference between the heating source and the temperature response,
\begin{equation} \label{tevolve}
T(\bm{r},t) = T_{0} + {H^{(ext)}_{0} \over c_{V} \omega}
\left[ \sin\omega t -  {q^{2} \over c_{V}}|\tilde{K}_{q}(\omega)| \sin{\left( \omega t - \delta \right)}
\right]\cos(\bm{q} \cdot \bm{r}) \text{   ,}
\end{equation}
in which $|\tilde{K}_{q}(\omega)|  = \sqrt{K_{q}^{\prime 2}(\omega) +K_{q}^{\prime \prime 2}(\omega) } $ is the magnitude of the response function and the phase angle is given by,
\begin{equation} \label{phase}
\tan \delta = {K_{q}^{\prime \prime } (\omega) \over K_{q}^{\prime } (\omega)} 
\end{equation}
Both the magnitude of the response function and the phase angle are therefore functions of the source frequency $\omega$.  It can be inferred that the oscillatory 
temperature tends to be damped most strongly when $K_{q}^{\prime } $ is large and $K_{q}^{\prime \prime } $ is relatively small, since then the response
is large and in-phase with the temperature fluctuations due to the source.

\begin{figure}
\begin{centering}
\includegraphics[scale=0.75]{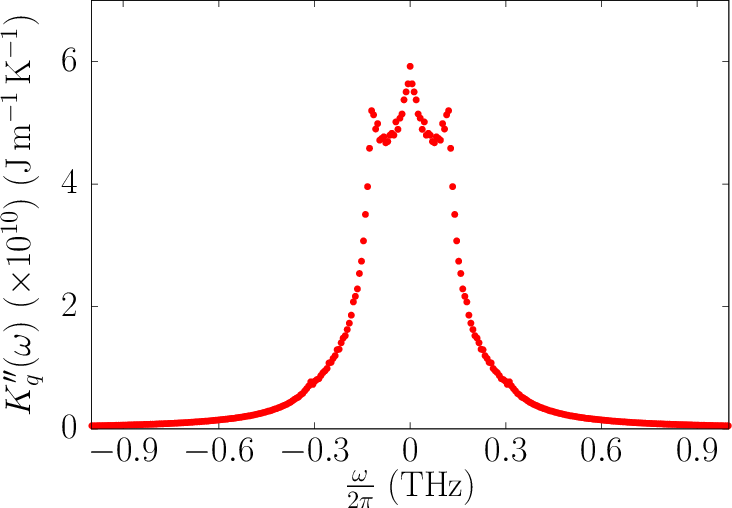} 
\caption{Magnitude of the response functions $| \tilde{K}_{q}\left(\omega\right)|$ for $q=q_{1}=
{2\pi \over L}$.
}
\label{fig18}
\end{centering}
\end{figure}

\begin{figure}
\begin{centering}
\includegraphics[scale=0.75]{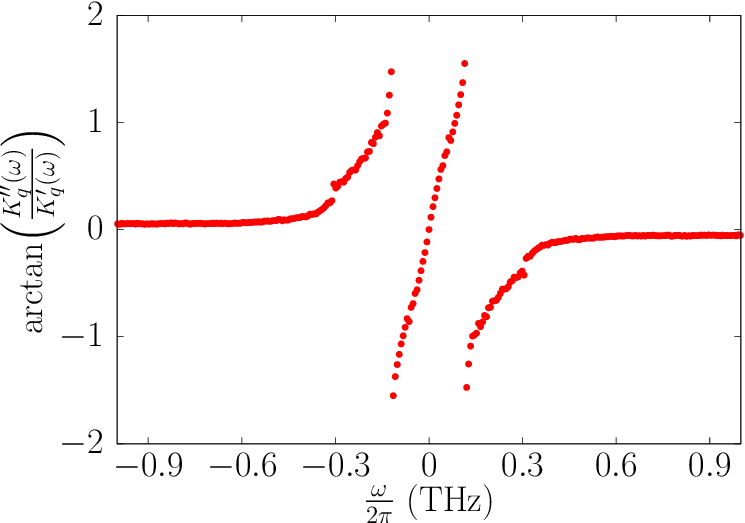} 
\caption{Phase angle $\delta=\arctan{\left(K_{q}^{\prime \prime} (\omega) \over K_{q}^{\prime}(\omega)\right)}$ for
$q=q_{1}={2 \pi \over L}$.
}
\label{fig19}
\end{centering}
\end{figure}

In Fig. \ref{fig18}, the magnitude $|\tilde{K}_{q}(\omega)|$ of the response function is plotted. In Fig. \ref{fig19}, the phase angle $\delta = \arctan{\left(K_{q}^{\prime \prime} \over K_{q}^{\prime}\right)}$ is shown as a function of frequency $\omega$. Not surprisingly, the magnitude is largest at the maximum of $K^{\prime}(\omega)$, which occurs at $\omega=0$, and also at the maxima
of  $K_{q}^{\prime \prime}(\omega)$. For very
small driving frequencies, the normal-modes are able to keep up with the perturbation, and dissipation of the excitation is maximized. Further away from $\omega=0$, the phase angle becomes an important factor, tending to admit larger-amplitude temperature deviations. For $\omega$ very near the peak of $K_{q}^{\prime \prime}(\omega)$, the phase angle tends to $\delta = \pm {\pi \over 2}$, and Eq. \ref{tevolve} shows that the temperature response is maximally out of phase with the source.  Finally, for frequencies outside of the range $\pm 0.3$THz, the normal modes are not able to keep up and are ineffective at dissipating the perturbation. This could have significant implications
for time-dependent sources, and moreover demonstrates how time-dependent sources could be used to probe the second-sound spectra to validate theoretical predictions.

For completeness, we also give the above evolution equation for $T(\bm{r},t)$ using the response function from the heat-diffusion equation,
\begin{equation}
T(\bm{r},t) = T_{0} + {H^{(ext)}_{0} \over  \kappa c_{V} } \left[
K_{q}^{\prime}(\omega) \cos \omega t + K_{q}^{\prime \prime}(\omega) \sin \omega t  
\right]\cos(\bm{q} \cdot \bm{r}) \text{  .}  
\end{equation}
This form appears different but in fact is equivalent to what was derived in Ref. \cite{Fernando_2020}. It is also interesting to note that the relationship between the $K_{q}^{\prime}(\omega)$ and
 $K_{q}^{\prime \prime}(\omega)$ terms and the $\sin \omega t$ and $\cos \omega t$ terms. This arises because in Fourier's law, the current response occurs
 immediately when the heat pulse is input, whereas the MD simulations and the computed response functions show that the current response lags behind the input heat pulse. 

\section{Discussion and Conclusions}

In this paper we have presented MD simulation results for thermal transport in single-layer graphene at length scales up to $\sim 68.1$nm and at $T=10$K at
scales $\sim 136.2$nm.
The results demonstrate oscillatory heat transport and strong deviations from Fourier's law at all simulated scales for $T=300$K.  For $T=10$K, sharp phonon lines associated with persistent resonance in the LA branch were also observed.  This is very consistent with our prior study of 
hBN where oscillatory transport was clearly observed at scales up to $110.8$nm for $T=100$K \cite{Schelling:2025aa}. In both of these systems, the limiting values of temperature and length scale
where oscillatory transport might be observed has not been fully established either theoretically or in experiment. Because we cannot establish the hydrodynamic regime, based on a restrictive definition of second sound, we can only hypothesize that our results might correspond to transport via second sound.

An important result is the prediction that for short TTG periods, the lifetime $\tau_{ss}$ is strongly dependent on the wave vector $\bm{q}$ of the temperature deviation. This is  in contradiction to
predictions based on most  BTE theories of second sound, which predict $\tau_{ss}$ is primarily dependent on anharmonic scattering rates and is independent on the vector $\bm{q}$ characterizing a perturbation. The exception to this appears to be the Green's function BTE theory developed in Ref. \cite{Chiloyan:2021uv} and to some extent recent BTE theory based on extension of the Guyer-Krumhansl approach \cite{Guyer:1966ab,Sendra:2021aa} which include nonlocal physics. Our predictions appear to be in qualitative agreement with TTG experimental results for graphite which show that $\tau_{ss}$ is shown to be strongly dependent on the grating period below $10 \mu$m \cite{huberman_2019,Ding_2022}. For very large grating periods ($> 10 \mu$m), $\tau_{ss}$ determined experimentally appears to saturate \cite{huberman_2019,Ding_2022}. In this limit, standard BTE theory is expected to be sufficient.

 As noted in our previous study of hBN, sharp spectral features may emerge in connection with resonant behavior in a particular band characterized by linear dispersion \cite{Schelling:2025aa}. In the MD simulations of graphene, very sharp features connected to resonance in the LA band were observed here in Fig. \ref{fig8} for a system at $T=10$K.  Sometimes in the literature these features are referred to as ``first sound''. However, in the
 physical picture developed here, these sharp features in the response function spectra $\tilde{K}_{q}(\omega)$ correspond to excitations which are not distinct from what we call ``second sound''. What might occur under certain conditions is that very specific phonon branches may be resonantly excited and propagate at a sufficient velocity to be detected independently from the ``thermal'' second sound. For example, in our hBN paper \cite{Schelling:2025aa}, even at the very high temperature $T=1200$K, distinct, coherent modes from the longitudinal-acoustic branch (LA) were observed in the calculations even when the second sound signal was essentially undetectable.
 
 Our theoretical understanding in Ref. \cite{Schelling:2025aa}, and also demonstrated here in the $\bm{q}$ dependence observed in Figs. \ref{fig5}-\ref{fig6}, indicates that as the period of a TTG experiment increases to larger length scales, the inherent linewidth due to the details of the phonon band structure becomes more narrow. This is because phonon resonances occur for frequencies $\omega \sim \bm{v}_{\bm{k}s} \cdot \bm{q}$. Hence, the dependence of $\tau_{ss}$ on the phonon band structure becomes less important as the period of a TTG experiment increases. At long enough grating periods, $\tau_{ss}$ will eventually be controlled almost entirely by phonon lifetimes, and hence standard BTE based models are sufficient for many experimental situations. This is because as the magnitude
 $|\bm{q}|$ decreases, the limiting bandwidth due to phonon resonances similarly decreases linearly with $|\bm{q}|$. Once the bandwidth due to phonon resonances becomes sufficiently narrow, the GK equation should reliably predict the observed $\tau_{ss}$.

The relatively weak oscillatory signal observed in our calculations, in contrast to experiments in graphite, most likely results primarily from two considerations. First, in graphite, it has been
observed that the parabolic ZA branch tends to become more linear in comparison to monolayer graphene. In the physical picture provided here, linear dispersion results in stronger resonance in oscillatory heat transport via the ZA branch, and hence a longer lifetime. Second, quantum statistics results in occupation of low-frequency acoustic modes, with higher-frequency optical modes unoccupied at low temperature. By contrast, our classical simulations result in contributions across the entire spectrum. Hence, classical MD simulations should result in a broader spectrum for $\tilde{K}_{\bm{q}}(\omega)$, and therefore a shorter lifetimes. However, our approximate analysis of this point apparently demonstrates only a small effect due to the use of classical statistics. More accurate predictions will emerge from theory that can account for quantum statistics. However, the approximate scaling of our MD results with weighting functions $W(\omega)$ did not seem to appreciably change the width of the spectral response function.
 
 The approach presented here to predict the response due to time-dependent sources has implications for experimental methods to elucidate
 second-sound spectra. Specifically, time-dependent experiments, possibly using TTG techniques, might be used to probe the spectral response. It was shown that both the magnitude of the response function and the phase angle between the source and response can be connected in a simple way to the observed temperature oscillations. This might be a way to directly probe second sound spectra to validate some of our predictions.

We also note that there have been time-dependent BTE methods developed which have similar objectives, namely to elucidate the response to time-dependent sources \cite{Hua:2020aa,Chiloyan:2021uv}. For example, the reported results for graphite were analyzed with this approach \cite{Chiloyan:2021uv}. It appears that there may be connections between the time-dependent BTE approaches and response functions. However, we have not yet tried to elucidate the relationships between these approaches. We do note that Ref. \cite{Chiloyan:2021uv} does lead to strong spectral response
at frequencies $\omega \sim \pm \bm{q} \cdot \bm{v}_{\bm{k},s}$, which is also present in the first-principles theory in our previous paper \cite{Schelling:2025aa} and described qualitatively in our analysis of the spectral response presented here. 

In our previous paper \cite{Schelling:2025aa}, we outlined how the response functions approach might be approached to be fully quantum-mechanical and use first-principles DFT calculations. The theory developed has connections with much earlier work by Sham
 \cite{Sham:1967aa,Sham:1967ab}. It is currently our objective to apply this methodology using the large dataset for graphene that we have used to compute thermal conductivity $\kappa$ using solutions to the LBTE. It is expected that correct quantum statistics will result in a more narrow spectral response, for the reasons argued above, and hence longer lifetimes $\tau_{ss}$.

 Finally, it has been brought to our attention the existence of two previous articles 
\cite{Michel:2017,Scuracchio:2019}that are closely related to our approach. It appears to us that the basic approach of those works is essentially equivalent to ours, with some distinctions. In particular, these previous works make connections to lattice displacement functions, and develop kinetic equations that connect more directly to phonon hydrodynamics. For example, Ref. \cite{Scuracchio:2019} obtains equations that describe the propagation of damped sound waves and local deviations of the phonon density distribution. Thus, it appears that the work in Refs. \cite{Michel:2017,Scuracchio:2019}, in contrast to our efforts, is more directly comparable and complementary to kinetic GK equation. We plan to further explore the connections between our work and these previous efforts. Also we note that in our previous work in Ref. \cite{Schelling:2025aa} we derived coupled equations based on solutions of the Bethe-Salpeter equation that are closely related to BTE methods. The connections between these expressions and those derived in Ref. \cite{Scuracchio:2019} for deviations from equilibrium are not yet entirely clear. We plan on establishing more clear connections to  \cite{Michel:2017,Scuracchio:2019} in part with the idea of developing our first-principles approach to better connect with existing theory to describe second sound and the hydrodynamic regime.
 
  \section{Acknowledgements}
  We acknowledge support from the NSF-ACCESS and the SDSC Expanse cluster for computing resources through Project PHY240153.
  We also acknowledge support and resources provided by UCF and the STOKES computer run by the UCF-ARCC administered by the Institute of Simulation and Training. The simulation data reported here was obtained using this two facilities. We also thank the referees for their constructive criticisms, and also the editor for directing our attention to Refs. \cite{Michel:2017,Scuracchio:2019} which represent an approach that is closely related to ours.
 
\newpage


\end{document}